\documentclass[journal]{IEEEtran}
\usepackage{cite}

\ifCLASSINFOpdf
  \usepackage[pdftex]{graphicx}
  \DeclareGraphicsExtensions{.pdf,.jpeg,.png}
\else
  \usepackage[dvips]{graphicx}
  \DeclareGraphicsExtensions{.eps}
\fi
\graphicspath{{Paper_results/TRACS/figs_v3/}{Paper_results/}}
\usepackage{amsmath}
\usepackage{amssymb}
\usepackage{booktabs}
\usepackage{multirow}
\usepackage[ruled,vlined]{algorithm2e}
\usepackage{amsthm}

\usepackage{stfloats}
\usepackage{color}

\begin{document}
%
\title{TRACS: A Geometry-Aware Framework for Scalable Multi-Agent Path Finding in Warehouses}
%
%
%

\author{Siddhant~Erande~and~Anuj~Tiwari%
\thanks{This work has been submitted to the IEEE for possible publication. Copyright may be transferred without notice, after which this version may no longer be accessible. \\
S.~Erande is with the Department of Engineering Design, and, A.~Tiwari is with Department of Mechanical Engineering at Indian Institute of Technology Madras, Chennai 600036, India.
E-mails: \texttt{siddhant.rs.1093@gmail.com}, \texttt{anujt@iitm.ac.in}}}

\maketitle

\begin{abstract}
Large-scale warehouse automation relies on efficient multi-agent path finding (MAPF) to coordinate thousands of robots in structured environments. Existing MAPF algorithms primarily improve conflict resolution while representing warehouses as generic navigation graphs, overlooking their inherent geometric structure and traffic patterns. This paper presents TRACS (Traffic-aware Routing and Aisle Coordination System), a geometry-aware planning framework that exploits warehouse layout to simplify planning rather than introducing another conflict-resolution algorithm. TRACS constructs a directed routing graph with alternating one-way aisles that eliminates head-on and edge-swap conflicts by design, decoupling spatial routing from temporal traffic coordination. Independent hybrid graph–grid routing is combined with lightweight edge-based scheduling to avoid joint space–time search while ensuring collision-free execution. Experimental evaluation on warehouse benchmarks against representative priority-based, iterative repair, and search-based MAPF planners shows that TRACS consistently achieves a 100\% empirical success rate while substantially improving planning scalability. On fixed-scene benchmarks with up to 1000 robots, TRACS reduces planning time by up to 14.7× while maintaining competitive makespan, lower flowtime, and near-optimal path quality. Under a fixed 10-minute planning budget, TRACS routes up to 5120 robots—roughly twice the largest fleet reached by the strongest baselines—while sustaining a 100\% success rate, demonstrating the effectiveness of exploiting warehouse geometry for scalable robotic warehouse systems.
 
\end{abstract}


\section{Introduction}

Autonomous Robotic Mobile Fulfillment Systems (RMFS) have become the backbone of modern warehouse automation, enabling hundreds to thousands of mobile robots to transport inventory between storage locations and workstations~\cite{wurman2008kiva}. As warehouse scale continues to increase, coordinating these robot fleets safely and efficiently has become a major challenge, making collision-free multi-robot path planning one of the principal computational bottlenecks in warehouse automation.

Multi-Agent Path Finding (MAPF) provides the standard framework for planning collision-free paths for multiple robots operating in a shared environment~\cite{stern2019mapf}. Over the past decade, considerable research has focused on improving MAPF scalability through increasingly sophisticated conflict-resolution strategies. Representative approaches include optimal search-based solvers such as conflict-based search (CBS)~\cite{sharon2015cbs}, increasing cost tree search (ICTS)~\cite{sharon2013icts}, M*~\cite{wagner2011mstar}, and SAT-based methods~\cite{surynek2016sat}; bounded-suboptimal and symmetry-breaking variants~\cite{li2021eecbs,li2019symmetry}; prioritized and coupled planning methods~\cite{erdmann1987multiple,silver2005cooperative,yu2013planning}; learning-based coordination~\cite{sartoretti2019primal}; iterative repair methods such as MAPF-LNS2~\cite{li2022lns2}; and scalable local-search planners including PIBT2~\cite{okumura2022pibt2} and LaCAM~\cite{okumura2023lacam}. Despite their methodological differences, these approaches share a common assumption, that is, the environment is represented as a generic navigation graph, while robot interactions are resolved dynamically in the joint space--time domain. Consequently, as robot density increases, an increasingly large fraction of computation is devoted to conflict resolution rather than geometric route planning.

Warehouse environments, however, exhibit considerably more structure than generic navigation domains. They consist of long parallel storage aisles connected through regular intersections, naturally producing predictable traffic patterns. Many industrial warehouse systems already employ directional traffic policies to reduce congestion and improve operational efficiency. Existing works have leveraged these characteristics through highway-guided search~\cite{cohen2016highways}, lifelong warehouse planning~\cite{ma2017lifelong,li2021lifelong}, and traffic-aware coordination. Nevertheless, these methods continue to operate on the original navigation graph and rely on online conflict resolution within the conventional MAPF framework.

In this work, rather than developing increasingly sophisticated mechanisms for resolving conflicts during planning, it is argued that warehouse structure should be incorporated directly into the planning representation. By embedding geometric properties of warehouse layouts into the routing graph, many conflicts can be eliminated before planning even begins, shifting part of the coordination burden from the planner to the environment itself. To realize this idea, this article proposes \textbf{TRACS (Traffic-aware Routing and Aisle Coordination System)}, a geometry-aware planning framework designed for structured warehouse environments. TRACS constructs a directed routing graph by assigning alternating travel directions to warehouse aisles, thereby eliminating head-on and edge-swap conflicts by construction while preserving global reachability. Route generation is decoupled from traffic coordination through a hybrid graph--grid planner that computes independent geometric routes, followed by a lightweight edge-based scheduler that resolves only the remaining local interactions. By preventing a large class of conflicts through representation rather than search, TRACS avoids expensive joint space--time exploration while ensuring collision-free execution. TRACS is intended for warehouse infrastructures where directional traffic policies can be naturally embedded into the routing graph. Within this setting, exploiting warehouse geometry simplifies planning while maintaining practical solution quality.

TRACS is evaluated through three complementary benchmark studies, described in Section~\ref{Section_exp_eval}, by quantifying, i) planning performance against representative search-based, priority-based and iterative-repair MAPF planners on fixed warehouse maps, ii) scalability under a fixed planning-time budget, and, iii) path quality relative to CBS. The primary contributions of this work are summarized as follows:




{
\begin{itemize}
    \item To embed the structural regularity of warehouse layouts into the planning representation through a geometry-aware directed routing graph that eliminates head-on and edge-swap conflicts by construction.

    \item To propose \textbf{TRACS}, a decoupled framework in which a hybrid graph--grid pipeline combining local grid planning with directed aisle navigation generates spatial routes, and a lightweight edge-based scheduler handles temporal coordination, in place of a further conflict-resolution algorithm.

    \item To experimentally demonstrate that exploiting warehouse geometry substantially improves planning scalability while maintaining competitive path quality across large-scale warehouse benchmarks.
\end{itemize}
}

\begin{figure*}[!tb]
\centering
\includegraphics[width=1.9\columnwidth]{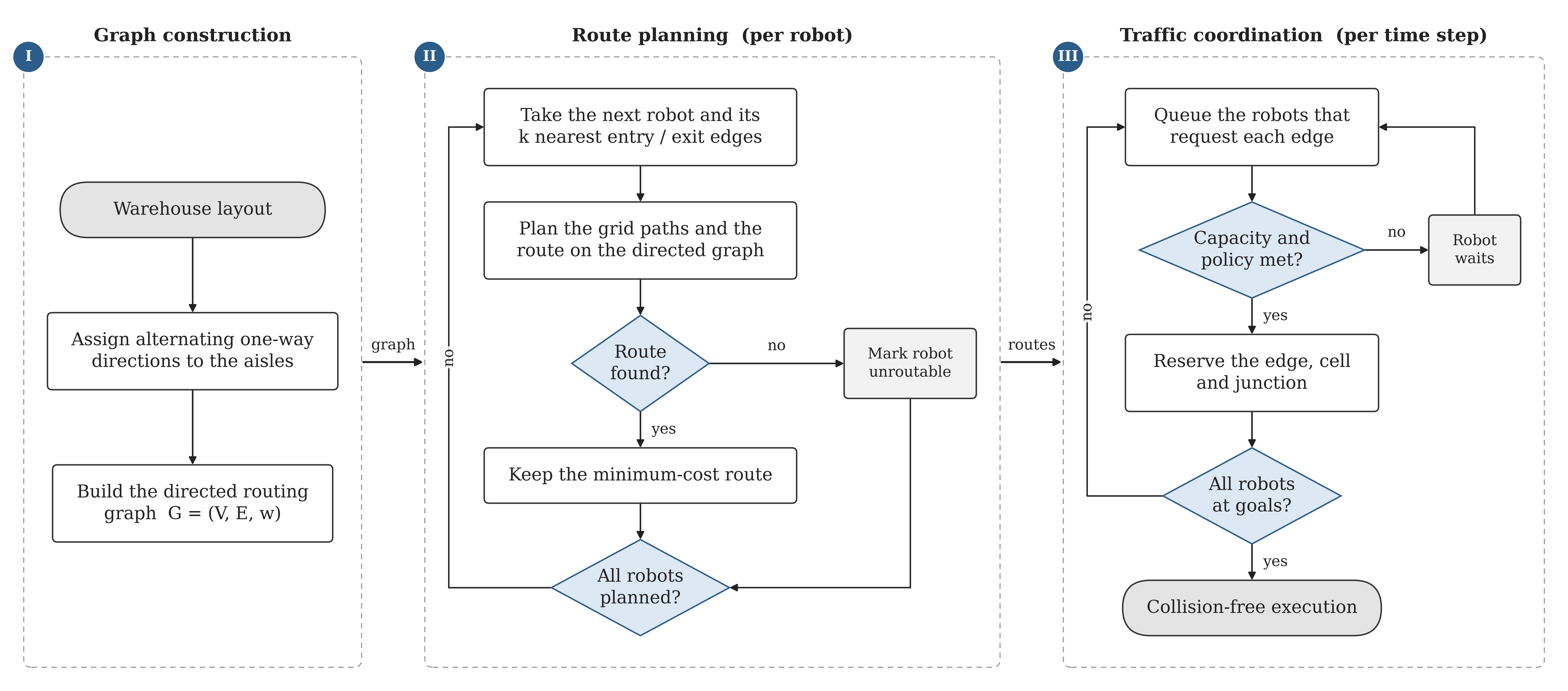}
\caption{Overview of the TRACS pipeline. (I) \emph{Graph construction}: the warehouse layout is converted into a directed routing graph by assigning alternating one-way travel directions to parallel aisles. (II) \emph{Route planning}: each robot independently computes its lowest-cost route on this graph, combining local grid search with directed-graph search over its $k$ nearest entry/exit edges. (III) \emph{Traffic coordination}: at execution time, a lightweight edge-based scheduler grants edge, cell, and junction reservations under capacity and headway constraints. Because stages II and III are decoupled, TRACS avoids joint space--time search while still guaranteeing collision-free execution.}
\label{fig:framework}
\end{figure*}
\section{Proposed Framework}

TRACS embeds the regular aisle structure of warehouses directly into the routing representation, so that major classes of conflicts are eliminated before planning begins and spatial routing can be decoupled from temporal coordination. 

{
As illustrated in Fig.~\ref{fig:framework}, the proposed framework consists of three stages: (I) the warehouse layout is converted into a geometry-aware directed routing graph that captures the warehouse topology together with directional traffic constraints; (II) each robot independently computes a spatial route using a hybrid graph--grid planner; and (III) an edge-based traffic scheduler coordinates execution through lightweight resource reservations and movement policies.}

\subsection{Geometry-Aware Warehouse Representation}

Warehouse layouts exhibit a highly regular topology, consisting of parallel storage aisles connected by perpendicular cross aisles. 
TRACS exploits this regularity by constructing a directed routing graph that carries traffic policies in the planning representation itself.

The warehouse is first modelled as a binary occupancy grid
$W=(C,O)$, where $C$ denotes the set of traversable cells and $O$ represents static obstacles and storage locations. From this grid, a directed routing graph
$G=(V,E,w)$ is constructed, where $V$ represents aisle intersections, $E$ denotes directed aisle segments, and
$w:E\rightarrow\mathbb{R}^{+}$ assigns each edge a traversal cost.

Unlike the occupancy grid, which contains every traversable cell, the routing graph retains only the topological decision points required for global navigation. This abstraction significantly reduces the search space while preserving connectivity between all reachable regions of the warehouse.

\subsubsection{Directed Aisle Assignment}

Let
$R=\{r_1,r_2,\ldots,r_m\}$
and
$C=\{c_1,c_2,\ldots,c_n\}$
denote the sets of horizontal and vertical aisles, respectively. Horizontal aisles are assigned alternating travel directions according to,

\begin{equation}
d(r_i)=
\begin{cases}
\rightarrow,& i\ \text{even},\\
\leftarrow,& i\ \text{odd},
\end{cases}
\end{equation}
\noindent
while vertical aisles follow

\begin{equation}
d(c_j)=
\begin{cases}
\uparrow,& j\ \text{even},\\
\downarrow,& j\ \text{odd}.
\end{cases}
\end{equation}
\noindent
Each directed edge is assigned a traversal cost proportional to its geometric length,
\begin{equation}
w(e)=\ell(e),
\end{equation}
\noindent
where $\ell(e)$ denotes the number of traversable cells belonging to edge $e$.

This alternating assignment preserves global reachability while preventing bidirectional travel within individual aisles. As a result, traffic is naturally distributed across neighboring corridors without requiring manually designed routing policies or additional traffic optimization.

\begin{figure}[!t]
\centering
\includegraphics[width=0.95\columnwidth]{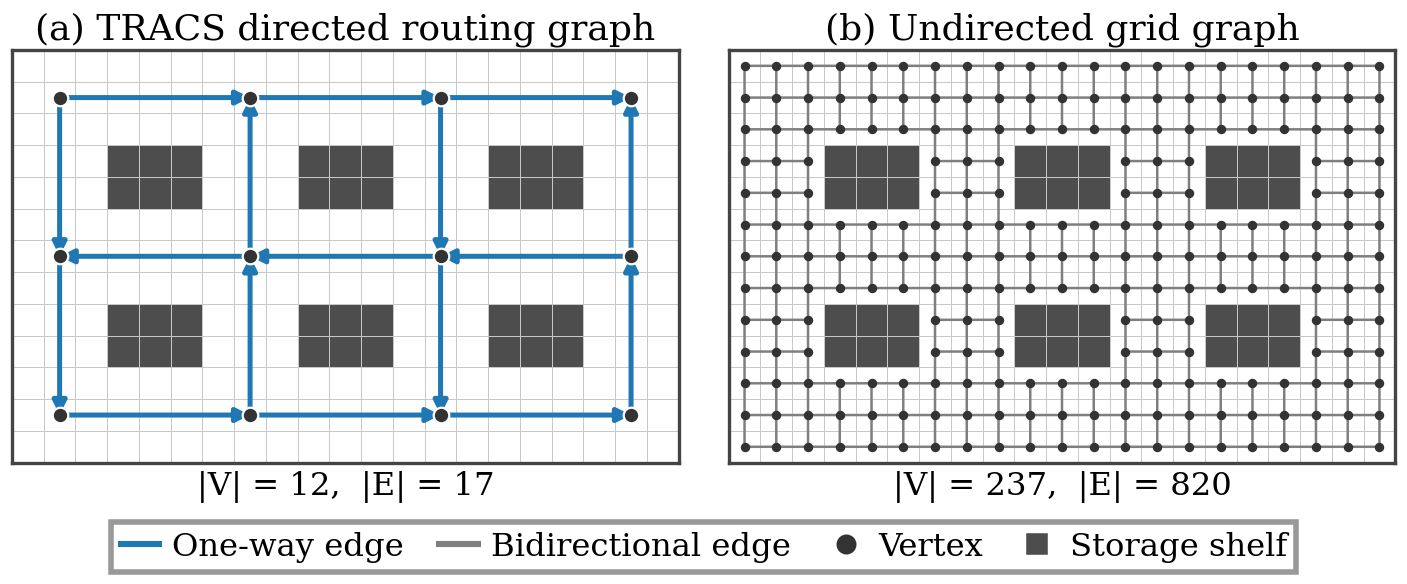}
\caption{Construction of the directed routing graph. Alternating aisle directions preserve global connectivity while eliminating head-on and edge-swap conflicts by construction. As a result, the remaining interactions are localized to junction nodes, which constitute only $1$--$2\%$ of the navigable cells in a typical warehouse.}
\label{fig:graph}
\end{figure}

\subsubsection{Conflict Elimination by Representation}

A key feature of TRACS is that many conflicts are removed through the graph construction itself rather than being resolved during planning.
\noindent
Head-on and edge-swap conflicts cannot occur along any directed aisle segment.
Each aisle segment is represented by a directed edge with a fixed traversal direction. Robots are therefore constrained to move only in the direction assigned to that edge. Since opposite-direction traversal is prohibited, two robots cannot simultaneously enter the same aisle segment from opposite ends. Consequently, both head-on and edge-swap conflicts are eliminated by construction. The only remaining interactions occur when multiple robots converge at shared junctions.

By eliminating these dominant classes of conflicts beforehand, the planning problem is fundamentally simplified. Instead of coordinating robot interactions throughout the entire warehouse, online coordination is restricted to the comparatively small number of junctions connecting directed aisle segments. Fig.~\ref{fig:graph} illustrates the construction of the directed routing graph for a representative warehouse layout. 

However, this simplification restricts the solution space. Because the alternating directed graph only contains a subset of the movements available on the occupancy grid, the construction preserves reachability but not optimality: routes that would traverse an aisle in an optimal solution are replaced by detours due to alternating directed edges. But we also show that in the (Fig.~\ref{fig:path_quality}) that beyond a certain heuristic separation between start and goal, the sub-optimality induced due to construction becomes lesser significant as opposed to start and goal points nearby. The construction also assumes parallel aisle groups; isolated corridors, dead-end aisles and single workstation access lanes cannot be made one-way without disconnecting the map, and are therefore retained as bidirectional edges that the scheduler treats as mutually exclusive resources. Since directions are fixed only at graph-construction time, demand-aware assignments can be introduced without altering the routing or scheduling stages.

\subsection{Hybrid Spatial Routing}

Once the directed routing graph has been constructed, each robot plans its route independently without considering the states of other robots. Consequently, route generation depends solely on the static warehouse geometry, while collision avoidance is handled later by the traffic coordination stage.

For a robot $r_i=(s_i,g_i)$ with start location $s_i$ and goal location $g_i$, the complete route $P_i$ is represented as a composition of three sub routes,
\begin{equation}
P_i=P_i^{entry}\oplus P_i^{graph}\oplus P_i^{exit},
\end{equation}
\noindent
where $P_i^{entry}$ sub route connects the start location $s_i$ to the routing graph (Fig.~\ref{fig:graph}~(a)) so the robot enters the directed routing, $P_i^{graph}$ is the sub route on the directed graph route to reach near the goal location $g_i$, and $P_i^{exit}$ connects the robot from the routing graph to the goal location $g_i$. The optimal route is selected by minimizing the total travel distance,
\begin{equation}
\begin{aligned}
P_i^{*}
=
\underset{
\substack{
P_i^{\mathrm{entry}},\,
P_i^{\mathrm{graph}},\,
P_i^{\mathrm{exit}}
}
}{\arg\min}
\Bigl(
L(P_i^{\mathrm{entry}})
+
L(P_i^{\mathrm{graph}})
+
L(P_i^{\mathrm{exit}})
\Bigr).
\end{aligned}
\end{equation}
where $L(\cdot)$ denotes the corresponding path length.

\subsubsection{Local Grid Routing}

Robot start and goal locations generally do not coincide with vertices of the routing graph. Therefore, local access paths are computed on the occupancy grid using A* search with the Manhattan-distance heuristic. Rather than restricting the search to the nearest aisle, the planner evaluates the $k$ nearest candidate entry and exit edges within a bounded search radius. This provides additional flexibility when one-way aisle constraints would otherwise introduce unnecessary detours.

\subsubsection{Global Graph Routing}

For every feasible entry--exit pair, the shortest path is computed on the directed routing graph using A* search. Edge costs are proportional to aisle lengths, while the heuristic is given by the Manhattan distance between graph vertices. To avoid repeated graph searches, computed routes are cached using the ordered entry--exit edge pair as the lookup key, allowing subsequent routing queries to be answered directly from memory.

\subsubsection{Candidate Route Selection}

Each candidate route is formed by concatenating the local entry path, the directed graph route, and the local exit path. The planner evaluates all feasible combinations and selects the entry--exit pair
$(e_{in}^{*},e_{out}^{*})$
that minimizes the total route cost,
\begin{equation}
L(P_i^{entry})+
L(P_i^{graph})+
L(P_i^{exit}).
\end{equation}

{Considering multiple nearby entry and exit edges significantly reduces the path overhead introduced by one-way aisle constraints while keeping the search space compact.}


\begin{algorithm}[t]
\caption{Hybrid Spatial Routing}
\label{alg:routing}

\KwIn{Warehouse grid, directed routing graph, start $s_i$, goal $g_i$}

\KwOut{Spatial route $P_i$}

Determine the $k$ nearest candidate entry and exit edges\;

$J^*\leftarrow\infty$\;

\ForEach{candidate entry edge}{

    Compute local A* path to the entry edge\;

    \ForEach{candidate exit edge}{

        Retrieve or compute the graph route\;

        Compute local A* path from the exit edge to the goal\;

        Evaluate the total route cost\;

        Update the best route if the cost decreases\;

    }

}

Return the minimum-cost composite route\;

\end{algorithm}


\subsection{Edge-Based Traffic Coordination}

The hybrid routing stage generates collision-unaware routes independently for every robot. Safe execution is ensured through an edge-based traffic coordination mechanism that regulates robot movement using lightweight local resource reservations. Since the directed routing graph eliminates head-on and edge-swap conflicts by construction, online coordination is required only for robots sharing aisle capacity or competing for the same junctions.


Each directed aisle segment is assigned a traversal capacity,
\begin{equation}
C(e)=\max\left(1,\left\lfloor\rho\,\ell(e)\right\rfloor\right),
\end{equation}
where $\rho\in(0,1]$ is a user-defined occupancy ratio and $\ell(e)$ denotes the geometric length of edge $e$. Consequently, longer aisle segments can safely accommodate multiple robots simultaneously while maintaining sufficient separation.

To coordinate robot movement, TRACS maintains four lightweight reservation structures,
\[
\mathcal{R}=\{L,R_c,R_v,H\},
\]
\noindent
where $L(e,t)$ records the occupancy of edge $e$ at time $t$, $R_c(c,t)$ stores cell reservations, $R_v(v,t)$ maintains junction reservations, and $H(e)$ records the latest entry time onto each edge.

Robots requesting access to the same edge are organized into edge-specific queues and admitted according to the available edge capacity while satisfying the movement policy. A robot is allowed to advance only if
\begin{equation}
\begin{aligned}
L(e,t) &< C(e), \qquad
R_c(c,t+1) = 0,\\
R_v(v,t+1) &= 0, \qquad
t-H(e) \ge \Delta.
\end{aligned}
\end{equation}
where the four conditions correspond respectively to available edge capacity, an unoccupied destination cell, an available destination junction when leaving an aisle, and a minimum temporal headway $\Delta$ between consecutive robots entering the same edge.

Once a robot is admitted, the edge occupancy, cell reservation, junction reservation, and headway timestamp are updated before the next scheduling decision. Since every scheduling decision depends only on local edge occupancy and neighboring resources, TRACS coordinates traffic without replanning complete trajectories or maintaining a global space--time reservation table.


\begin{algorithm}[t]
\caption{Edge-Based Traffic Coordination}
\label{alg:scheduler}

\KwIn{Planned graph routes}

\KwOut{Collision-free execution}

Initialise reservation tables and edge queues\;

\While{unfinished robots remain}{

Insert robots requesting their next edge into the corresponding queues\;

\ForEach{edge with waiting robots}{

\While{edge capacity is available}{

Select the next robot from the queue\;

\eIf{movement policy is satisfied}{

Reserve edge, cell, and junction resources\;

Update headway information and robot state\;

}{
Robot waits until the next scheduling cycle\;
}

}

}

Advance the simulation time\;

}

\end{algorithm}


\subsection{Computational Complexity}

The computational cost of TRACS consists of three components, graph construction, per-robot spatial routing, and online traffic coordination.

\noindent
{\it 1. Graph construction:} The directed routing graph is constructed once during preprocessing by traversing the occupancy grid, requiring
$O(HW)$
time for a warehouse of dimensions
$H\times W$. This preprocessing cost is independent of the number of robots.

\noindent
{\it 2. Per-robot spatial routing cost:} During route generation, each of the $N$ robots evaluates at most $k$ candidate entry and $k$ candidate exit edges, resulting in at most $k^2$ candidate routes. Each candidate requires two local A* searches on the occupancy grid and one A* search on the routing graph, giving a total routing complexity of
\noindent
\begin{equation}
O\!\left(
Nk^2
\left(
HW\log(HW)
+
|E|\log|V|
\right)
\right),
\end{equation}
\noindent
where $|V|$ and $|E|$ denote the numbers of graph vertices and directed edges, respectively. In practice, caching graph routes using ordered entry--exit edge pairs substantially reduces repeated graph searches.

\noindent
{\it 3. Online coordination:} During execution, each robot is processed at most once per scheduling cycle. Since reservation updates, queue operations, occupancy checks, and movement-policy evaluations all require constant time, the scheduling complexity over a planning horizon of $T$ time steps is
\noindent
\begin{equation}
O(NT).
\end{equation}
\noindent
Combining all three stages yields an overall computational complexity of
\noindent
\begin{equation}
O\!\left(
HW+
Nk^2
\left(
HW\log(HW)
+
|E|\log|V|
\right)
+
NT
\right).
\end{equation}
\noindent
Because routing is performed independently for each robot and online coordination is limited to lightweight local resource management, the dominant computational cost arises from route generation rather than conflict resolution. This enables TRACS to scale efficiently to large robot fleets operating in structured warehouse environments.
\section{Experimental Evaluation}
\label{Section_exp_eval}
\subsection{Experimental Setup}

We evaluate TRACS on representative warehouse environments and compare its performance against several state-of-the-art MAPF planners, including Prioritized Planning, PIBT2, LaCAM, and MAPF-LNS2. All methods are executed on identical warehouse layouts and robot configurations to ensure a fair and consistent comparison. All experiments are conducted on a machine equipped with an Intel Core i9-14900 CPU and 64~GB of RAM, running Ubuntu 22.04 LTS.

The evaluation follows two complementary experimental protocols. The first is a \emph{fixed-scene} study performed on a common warehouse layout (constructed as shown in Fig.~\ref{fig:graph}), where the number of simultaneously active robots is varied from 100 to 1000. This experiment measures planning time and solution quality as the fleet size increases.

The second protocol is an \emph{incremental-budget} stress test conducted on two larger warehouse layouts with different dimensions and aisle spacings ($207\times457$ and $307\times927$). Starting with 10 robots, the fleet size is doubled after each successful run. The process continues until a planner is unable to compute a solution within a 10-minute time limit, at which point the largest fleet successfully planned within the allotted budget is recorded.

Our evaluation considers four complementary performance metrics, \\
\noindent 
\textit{i) success rate}, which is the percentage of robots that successfully reach their assigned goals, \\
\noindent \textit{ii) planning time}, which is the time required to compute feasible routes, excluding graph preprocessing. \\
\noindent \textit{iii) path quality}, which is the ratio between the executed path length and the corresponding shortest-path length, and, \\
\noindent \textit{iv) traffic efficiency}, which is measured through average edge occupancy and the spatial distribution of traffic throughout the warehouse.

All competing planners are evaluated on the same benchmark instances using identical start-goal assignments, and TRACS uses the same directed routing graph and a fixed edge occupancy ratio $\rho$ across all experiments.
\subsection{Scalability Analysis}

\begin{figure*}[!tb]
\centering
\includegraphics[width= 1.95\columnwidth]{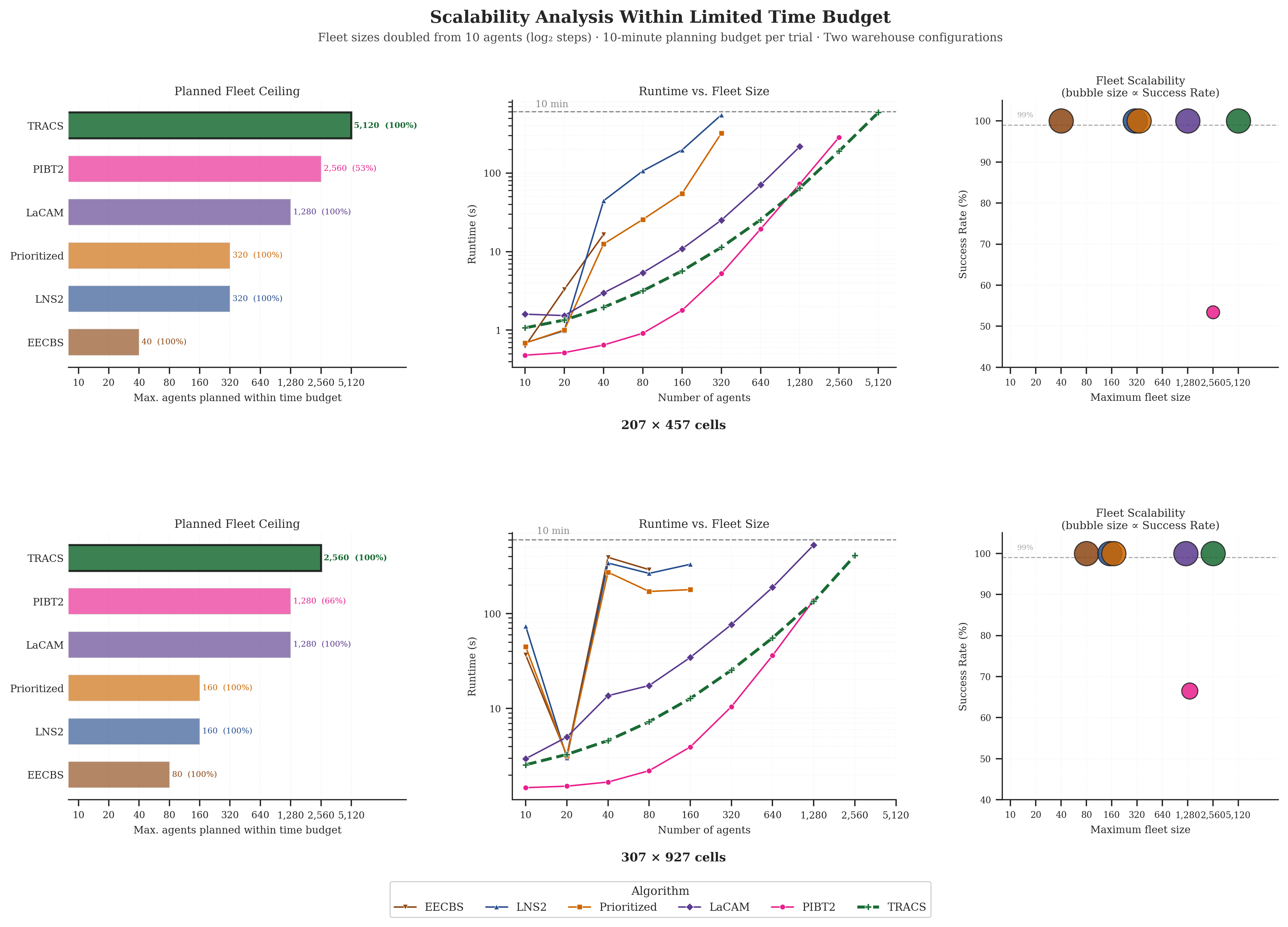}
\caption{Scalability under a fixed 10-minute planning budget on two large warehouses ($207\times457$, top; $307\times927$, bottom). \textit{Left}: largest fleet each planner completes within budget. \textit{Centre}: planning time versus fleet size (log--log scale), with the 10-minute ceiling marked. \textit{Right}: success rate versus maximum fleet size (bubble size $\propto$ success rate). TRACS reaches the largest fleet size on both maps while remaining on the $100\%$ success-rate line throughout.}

\label{fig:runtime}
\end{figure*}

The primary objective of TRACS is to improve planning scalability by exploiting the inherent structure of warehouse environments rather than relying on increasingly sophisticated conflict-resolution algorithms. Fig.~\ref{fig:runtime} compares the planning performance of TRACS against representative MAPF planners as the number of robots increases on the fixed-scene benchmark.

Across all fleet sizes, TRACS achieves a $100\%$ empirical success rate while consistently requiring less planning time than the competing methods. Since robots compute their routes independently and bidirectional aisle conflicts are eliminated during graph construction, online coordination is limited to local contention at shared resources instead of global space--time conflict resolution. Consequently, the planning cost increases much more gradually as the fleet grows, and the performance gap becomes increasingly pronounced at high robot densities where conflict resolution dominates the computation of conventional MAPF planners.

Despite the directional routing constraints, TRACS maintains a makespan that remains competitive with existing approaches (Fig.~\ref{fig:runtime}, right). At $N=1000$ on the $81\times125$ warehouse (Table~\ref{tab:comparison}), TRACS achieves the fastest planning time among all planners with a $100\%$ success rate, requiring only $42.7$~s. This corresponds to speedups of $14.7\times$ over Prioritized Planning and $10.3\times$ over MAPF-LNS2. In addition, TRACS produces the lowest overall flowtime, indicating that improved scalability is achieved without sacrificing solution quality. 

\begin{table}[!t]
\centering
\caption{Planner comparison at $N=1000$ on the $81\times125$ warehouse. Among planners achieving a $100\%$ success rate (SR), TRACS provides the fastest planning time and the lowest flowtime. Times are reported in seconds.}
\label{tab:comparison}
\setlength{\tabcolsep}{4.5pt}
\begin{tabular}{lcccc}
\toprule
Planner & SR & Time (s) & Makespan & Flowtime \\
\midrule
PIBT2~\cite{okumura2022pibt2}          & 67\%           & 93.5          & 240 & 240{,}000 \\
Prioritized~\cite{erdmann1987multiple} & 100\%          & 629.8         & 177 & 98{,}270 \\
MAPF-LNS2~\cite{li2022lns2}            & 100\%          & 439.3         & 177 & 98{,}270 \\
LaCAM~\cite{okumura2023lacam}          & 100\%          & 77.5          & 178 & 178{,}000 \\
\textbf{TRACS (ours)}                  & \textbf{100\%} & \textbf{42.7} & 182 & \textbf{73{,}635} \\
\bottomrule
\end{tabular}
\end{table}

\subsubsection{Maximum Fleet Size Under a Fixed Planning Budget}

While the previous experiment evaluates scalability under increasing robot populations, a complementary question is how large a fleet each planner can successfully route within a practical planning budget. To answer this, we evaluate all methods on two larger warehouse layouts using a fixed time limit of 10 minutes.

As summarized in Table~\ref{tab:budget} and Fig.~\ref{fig:runtime}, TRACS successfully plans routes for up to $5120$ robots on the $207\times457$ warehouse and $2560$ robots on the larger $307\times927$ layout. These limits are approximately twice those achieved by the strongest baseline planners (PIBT2 and LaCAM) and nearly an order of magnitude larger than those supported by Prioritized Planning and MAPF-LNS2. EECBS, the bounded-suboptimal search baseline, exhausts the time budget far earlier (at $40$--$80$ robots), highlighting the difficulty of maintaining near-optimal solutions through search at scale. TRACS instead trades strict optimality for scalability through its directed routing structure; however, as shown in Sec.~\ref{sec:path_quality}, this tradeoff remains small, with route lengths staying within a few percent of the optimal CBS reference for longer trips. Importantly, TRACS maintains a $100\%$ empirical success rate across all successful runs (Fig.~\ref{fig:runtime}, right column), whereas PIBT2's larger fleet ceiling on the $207\times457$ map is achieved at a reduced success rate of only $53\%$.


As in the fixed-scene study, the planning budget is spent generating routes rather than resolving interactions between robots. Consequently, the maximum fleet size that can be accommodated continues to grow with the available computation time instead of saturating under increasing traffic density.

\begin{table}[!t]
\centering
\caption{Largest fleet successfully routed within a 10-minute planning budget. Fleet size is doubled from 10 robots until the first unsuccessful run.}
\label{tab:budget}
\begin{tabular}{lcc}
\toprule
Planner & $207\times457$ & $307\times927$ \\
\midrule
Prioritized~\cite{erdmann1987multiple} & 320 & 160 \\
MAPF-LNS2~\cite{li2022lns2}            & 320 & 160 \\
LaCAM~\cite{okumura2023lacam}          & 1280 & 1280 \\
PIBT2~\cite{okumura2022pibt2}          & 2560 & 1280 \\
\textbf{TRACS (ours)}                  & \textbf{5120} & \textbf{2560} \\
\bottomrule
\end{tabular}
\end{table}
\subsection{Path Quality}
\label{sec:path_quality}

\begin{figure}[!t]
\centering
\includegraphics[width=\columnwidth]{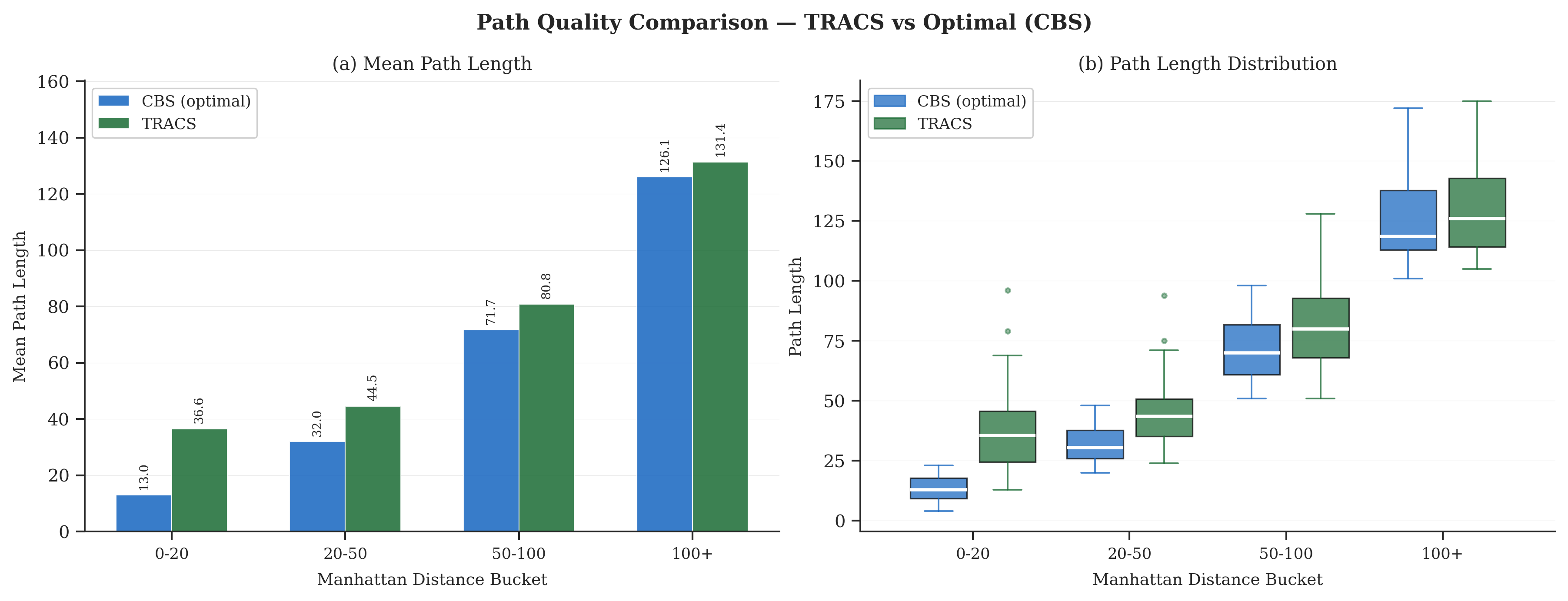}
\caption{Path length of TRACS versus the single-agent-optimal CBS baseline, grouped by Manhattan trip distance: (a) mean path length; (b) full distribution (box: interquartile range; whiskers: 5th--95th percentile). The one-way detour overhead is largest, in relative terms, for the shortest trips ($0$--$20$ cells) and shrinks to within a few percent of optimal beyond $\sim\!20$ cells, where warehouse shelf-to-workstation trips predominate.}

\label{fig:path_quality}
\end{figure}

The use of a directed routing graph introduces one-way traffic constraints, which can increase travel distance relative to unconstrained shortest paths, particularly for nearby destinations. To quantify this effect, we compare the route lengths generated by TRACS against those obtained using CBS, which produces single-agent optimal paths, on the $81\times125$ warehouse layout. Randomly sampled start--goal pairs are grouped according to their shortest-path distance into four buckets ($0$--$20$, $20$--$50$, $50$--$100$, and $100+$ cells), and the path-length overhead is measured over repeated trials (Fig.~\ref{fig:path_quality}).

As shown in Fig.~\ref{fig:path_quality}, the additional travel distance is concentrated almost entirely among the shortest trips. For paths shorter than $20$ cells, a single detour introduced by the one-way aisle structure represents a relatively large proportion of the overall travel distance, resulting in the highest median overhead. Beyond this range, however, the overhead rapidly decreases. For trips longer than approximately $20$ cells, the median path length remains within only a few percent of the shortest-path optimum, while for the longest distance buckets it is occasionally slightly lower than the CBS baseline because CBS optimizes jointly feasible solutions rather than individual path lengths.

This behavior is expected. As travel distance increases, robots have access to a larger number of feasible aisle combinations, allowing the planner to recover routes that are effectively near-optimal despite the directional constraints. 
Since warehouse operations are predominantly composed of longer shelf-to-workstation transport tasks, the modest overhead observed for very short trips has only a limited impact on overall system performance while enabling substantially improved planning scalability.

\begin{figure*}[!th]
\centering
\includegraphics[width=1.8\columnwidth]{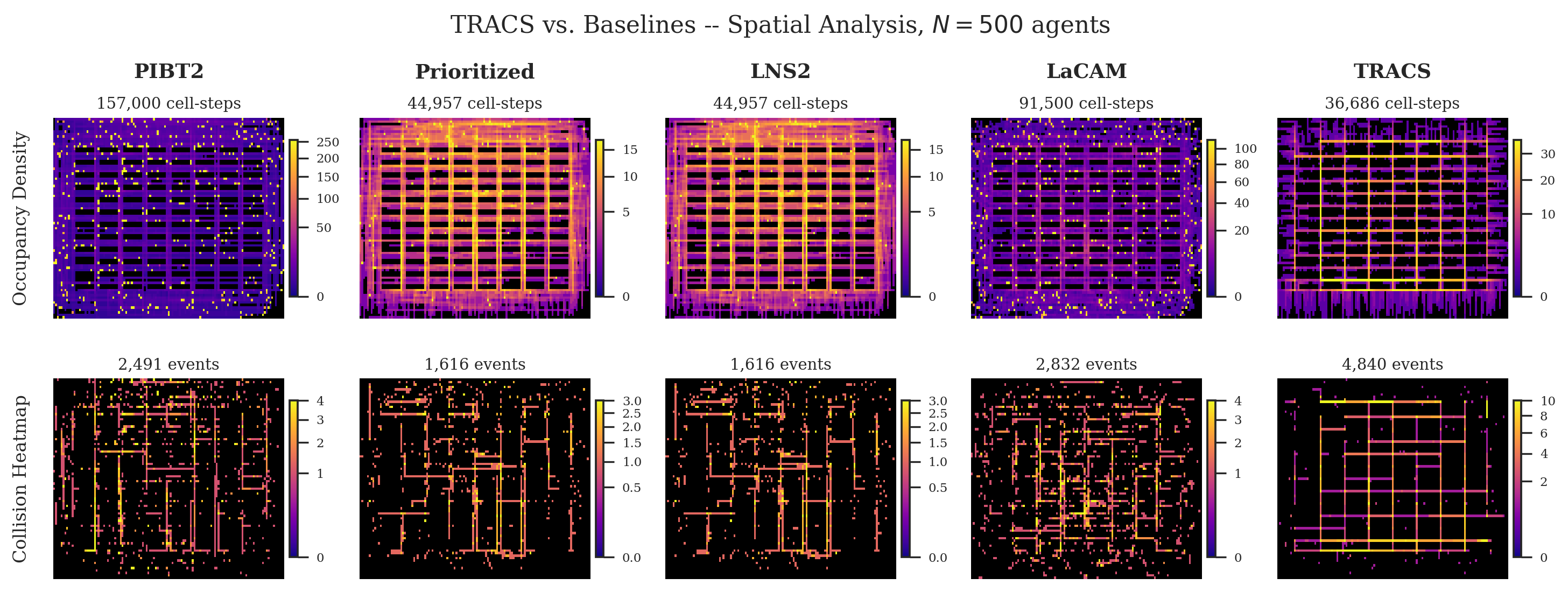}
\caption{Occupancy density (top row) and collision-event density (bottom row) for $N=500$ robots, compared across all five planners. TRACS spreads traffic along the directed aisle backbone and visits the fewest total cells (top row), while its higher count of local following/goal-contention events (bottom row) reflects the edge-based scheduler actively managing junction access rather than unresolved conflicts.}
\label{fig:heatmap}
\end{figure*}

\subsection{Traffic Analysis}

To better understand the impact of the proposed routing representation, we analyze the resulting traffic and contention patterns generated during execution. Fig.~\ref{fig:heatmap} compares occupancy density and collision-event density across all five planners for $N=500$ robots.

As shown in Fig.~\ref{fig:heatmap}, TRACS distributes traffic evenly along the directed aisle backbone and visits substantially fewer cells overall than every baseline ($36{,}686$ cell-visits, versus $44{,}957$--$157{,}000$), confirming that the alternating one-way assignment removes redundant back-and-forth travel. The collision-event heatmap shows the opposite ranking: TRACS records \emph{more} local events than the baselines, not fewer. This is expected rather than a weakness, because head-on and edge-swap conflicts are already eliminated by construction, so essentially all recorded events for TRACS are following and goal-contention interactions generated precisely by the edge occupancy limits, junction reservations, and temporal headway constraints that the scheduler uses to keep execution collision-free. This confirms that the proposed representation shifts coordination away from expensive global path planning and onto lightweight, localized traffic management, trading a higher rate of cheap local queueing for markedly faster planning and lower cell occupancy.

\begin{figure*}[!t]
\centering
\includegraphics[width=1.8\columnwidth]{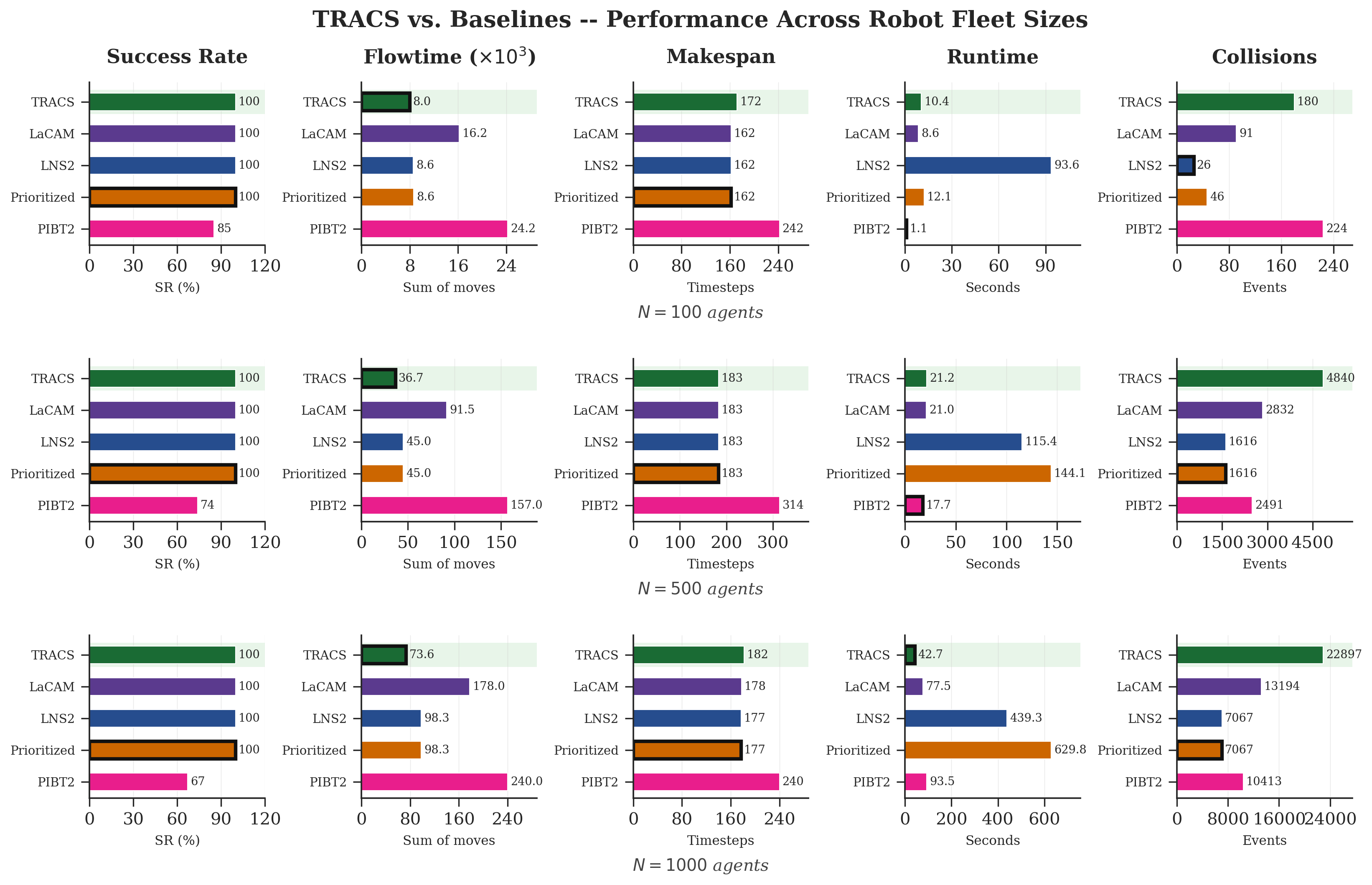}
\caption{Success rate, flowtime, makespan, planning time, and local contention events for TRACS and all baselines at $N=100$, $500$, and $1000$ robots on the $81\times125$ warehouse. TRACS is the only planner with a $100\%$ success rate at every fleet size, and its advantage in planning time and flowtime widens as $N$ grows.}
\label{fig:occupancy}
\end{figure*}

Fig.~\ref{fig:occupancy} consolidates all five metrics (success rate, flowtime, makespan, planning time, and local contention events) across the three fixed-scene fleet sizes ($N=100$, $500$, $1000$), making the trend as robot density increases directly comparable across planners. TRACS is the only planner that sustains a $100\%$ success rate at every fleet size; PIBT2, the closest competitor on raw planning time, degrades from $85\%$ at $N=100$ to $67\%$ at $N=1000$. TRACS also keeps the lowest flowtime at every $N$, and its planning time grows the most gradually of all planners as $N$ increases, widening the gap over the search- and repair-based baselines at higher density. The local contention-event count for TRACS grows faster than for the baselines, which is consistent with Fig.~\ref{fig:heatmap}: these are junction-queueing and following events handled by the edge-based scheduler rather than unresolved collisions, and their growth reflects increasing aisle utilization rather than a degradation in solution quality, since success rate and flowtime both continue to improve relative to the baselines as $N$ grows.

\section{Conclusion}


This paper introduced TRACS, a geometry-aware framework for scalable multi-agent path planning in structured warehouse environments. Instead of improving conflict resolution within the conventional MAPF formulation, TRACS embeds directional traffic policies into the routing representation, which eliminates head-on and edge-swap conflicts by construction and allows hybrid graph--grid routing to be decoupled from lightweight edge-based scheduling. On large-scale warehouse benchmarks the framework consistently achieves a $100\%$ empirical success rate while significantly improving planning scalability over representative MAPF planners. Directional routing introduces a modest path-length overhead for very short trips, but this overhead rapidly diminishes with travel distance, making the approach well suited to the long-distance transport tasks that dominate robotic warehouse operations.

More broadly, this work shows that exploiting domain structure through the planning representation is an effective alternative to increasingly sophisticated conflict-resolution algorithms.
Future work will extend TRACS to dynamic task allocation, heterogeneous robot fleets, adaptive traffic policies and replanning under edge failure, and deployment in real-world warehouse environments with dynamic obstacles and execution uncertainties.

%



\end{document}